\documentclass[aps,prl,twocolumn,showpacs,superscriptaddress]{revtex4}

\usepackage{graphicx}
\usepackage{dcolumn}
\usepackage{amsmath}
\usepackage{enumerate,amsthm,amssymb,color}
\usepackage{bbm}
\usepackage{bm}
\usepackage[french, english]{babel}

\newcommand{\COMMENT}[1]{}

\begin{document}

\preprint{APS/123-QED}

\title{Practical Quantum Coin Flipping}

\author{Anna Pappa}\email{anna.pappa@telecom-paristech.fr}\affiliation{LTCI, CNRS - T\'el\'ecom ParisTech, Paris, France}
\author{Andr\'e Chaillloux}\affiliation{LIAFA, CNRS - Universit\'e Paris 7, Paris, France}
\author{Eleni Diamanti}\affiliation{LTCI, CNRS - T\'el\'ecom ParisTech, Paris, France}
\author{Iordanis Kerenidis}\affiliation{LIAFA, CNRS - Universit\'e Paris 7, Paris, France}

\date{\today}

\begin{abstract}
We show that a single quantum coin flip with security guarantees that are strictly better than in any classical protocol is possible to implement with current technology. Our protocol takes into account all aspects of an experimental implementation like losses, multi-photon pulses emitted by practical photon sources, channel noise, detector dark counts and finite quantum efficiency. We calculate the abort probability when both players are honest, as well as the probability of one player forcing his desired outcome. For channel length up to 21 km, we achieve honest abort and cheating probability that are better than in any classical protocol. Our protocol is easy to implement using attenuated laser pulses, with no need for entangled photons or any other specific resources.
\end{abstract}

\maketitle
{\bf Introduction} -- Coin Flipping is a fundamental cryptographic primitive with numerous applications, where two distrustful parties separated by distance wish to agree on a random bit \cite{blum}. Classically, it is impossible to have a coin-flipping protocol with cheating probability less than 1, unless computational assumptions are considered. In other words, a dishonest player can force the outcome of the coin flip with probability 1. In the quantum model, where the two parties share a quantum channel, it was also proven that perfect coin flipping is information-theoretically impossible~\cite{lochau98}. On the other hand, several quantum protocols have been proposed that achieve a cheating probability lower than 1~\cite{ahar,ambainis,kerchai}. 

These results are important from a theoretical point of view, however they assume perfect implementation of the protocols. The situation is more subtle when we deal with realistic conditions encountered in experimental implementations,  for example multi-photon pulses emitted by practical sources, losses and channel noise, since if taken into account they may render the protocols insecure in practice ~\cite{ahar,ambainis,kerchai,exp,nguyen}. 

To study the security of practical coin flipping protocols, in addition to the cheating probability, we need to take into account the probability that the honest players abort the protocol. H\"anggi and Wullschleger \cite{han} gave tight bounds for the cheating probability $p$ of any classical or quantum coin flipping protocol, where the honest players abort with probability $H$. 
Hence, the goal of any quantum implementation of coin flipping would be to achieve both honest abort and cheating probability strictly smaller than in any classical protocol.

Recently, protocols that address some of the issues that arise in experimental implementations have been proposed. 
A major step was taken by Berlin et al~\cite{berlin}, who proposed a protocol that is completely impervious to losses and achieves cheating probability 0.9. In their theoretical analysis, they do not deal with noise and thus the honest abort probability is always zero. The main disadvantage of this protocol is that it becomes completely insecure in the presence of multi-photon pulses, i.e. when the implementation is based on an attenuated laser source rather than a perfect single-photon source.  Subsequently, Berlin et al \cite{flip} implemented this protocol using a source of entangled qubits and hence avoiding multi-photon pulses. In order to deal with the noise present in the experiment, they defined a different primitive, {\em sequential coin flipping}, instead of a single coin flip, and conjectured that this primitive still remains impossible classically. 

Chailloux~\cite{chai10} proposed an improved protocol with cheating probability 0.86. Last, Barrett and Massar \cite{massar} had introduced the primitive of string coin flipping, as an alternative to what they considered impossible to achieve experimentally, i.e. a single coin flip. However, this primitive turned out to be possible classically.

\COMMENT{
Over the last years, there have been some implementations of quantum coin flipping protocols that attempted to deal with the practical conditions. In \cite{exp}, the authors implement Ambainis' protocol \cite{ambainis} using entangled qutrits, a protocol which is proven to be insecure in the presence of channel or detector losses. Moreover, they perform their experiment at visible wavelength, which is not suitable for long-distance usage over telecommunication networks. Berlin et al \cite{flip} present the first implementation of a loss-tolerant quantum coin flip, which nevertheless is noise-intolerant. Finally, Nguyen et al \cite{nguyen} present an implementation that also takes into account some experimental imperfections.
}

{\bf Our work} -- In the current letter, we present a quantum coin flipping protocol that can be implemented using today's technology and that has both honest abort and cheating probability provably lower than the ones achieved by any classical protocol \cite{han}. Our protocol uses a standard attenuated laser source and we analyse how all practical aspects, like multi-photon pulses, channel noise, system loss, detector dark counts and finite quantum efficiency, affect the honest abort and cheating probability. We prove that for a single coin flip, if the noise is up to $2\%$ and for channel length up to $21$km, we can achieve at the same time honest abort probability ($\approx 1\%$) and cheating probability ($\approx 0.91$) strictly smaller than classically possible. We note that, similar to quantum key distribution protocols \cite{QKD}, our protocol is not completely impervious to losses, but can tolerate up to a certain amount of losses, which corresponds to distances in typical metropolitan area networks.

{\bf Protocol} -- Our protocol is a refinement of the one proposed by Berlin et al~\cite{berlin}; the main difference is that Alice sends $K$ pulses instead of one, and uses an attenuated laser source to produce her states, instead of a perfect single-photon or an entangled-photon source:
\begin{enumerate}
\item Alice sends $K$ photon pulses to Bob, where the number of photons in each pulse $i$ follows the Poisson distribution with $p_i=e^{-\mu}\mu^i/i!$ and mean photon number $\mu$. She prepares each pulse in the state $\lvert\phi_{\alpha_i,c_i}\rangle$, $i=1,...,K$, such that:
\begin{eqnarray*}
\lvert\phi_{\alpha_i,0}\rangle&=&\sqrt{a}\lvert0\rangle+(-1)^{\alpha_i}\sqrt{1-a}\lvert1\rangle\\
\lvert\phi_{\alpha_i,1}\rangle&=&\sqrt{1-a}\lvert0\rangle-(-1)^{\alpha_i}\sqrt{a}\lvert1\rangle
\end{eqnarray*}
where $\alpha_i\in_R\{0,1\}$ is the basis and $c_i\in_R\{0,1\}$ is the bit chosen by Alice.
\item Bob picks a measurement basis $\alpha'_i$ for every pulse. If his detectors do not click for any pulse, then he aborts. Else, let $j$ the first pulse he detected.
\item Bob picks $c'_j\in_R\{0,1\}$ and sends it to Alice, together with the index $j$.
\item Alice reveals $\alpha_j,c_j$.
\item If $\alpha_j=\alpha'_j$, Bob checks that the outcome of his measurement is indeed $\lvert\phi_{\alpha_j,c_j}\rangle$, otherwise he aborts.
\item If Bob has not aborted, then the outcome of the protocol is $b=c_j+c'_j$.
\end{enumerate}

{\bf Honest Player Abort} -- Any amount of noise in an experimental implementation results in a non-zero honest abort probability. Here, we analyse how exactly noise and the other experimental parameters affect the honest abort probability in order to ensure that the protocol achieves a task which remains impossible classically. We note that a similar analysis can also be done for the Berlin et al. protocol. 
The situations where an honest abort might occur with some probability are the following:
\begin{enumerate}
\item Bob's detectors do not click in any of the $K$ rounds of the coin flip. The abort probability is 1.
\item Bob's first detection is due to a dark count. The abort probability is 1/4, since if $\alpha_j=\alpha'_j$ (step 5), he will abort with probability 1/2 (dark count is totally random), else if  $\alpha_j\neq\alpha'_j$ he will not abort.
\item The noise in the channel alters the state of the photon. The abort probability is 1/2, since he will only abort if $\alpha_j=\alpha'_j$ (step 5).
\end{enumerate}
The total honest abort probability is then:
\begin{align*}
H &=\textstyle{Z^K(1-d_B)^K    +\frac{1}{4}\sum_{i=1}^K (1-d_B)^{i-1} d_B  Z^i} \\
&+\textstyle{\big[1-Z^K(1-d_B)^K-\sum_{i=1}^K (1-d_B)^{i-1} d_B  Z^i\big]\frac{e}{2}}
\end{align*}
where $Z = p_0+\big(1-p_0\big) \big(1-F\eta \big)$: probability that no signal arrives at Bob's detectors; $F$: system transmission efficiency; $\eta$: detector finite quantum efficiency; $d_B$: probability of detector dark count; $e$: probability of wrong measurement outcome due to noise.

{\bf Malicious Alice} -- Alice's optimal cheating strategy in our protocol is the same as the one in the Berlin et al's protocol. We assume Alice to be all-powerful, which means that she controls all aspects of the implementation, including the errors in Bob's detectors. 
It is in her best interest to replace the lossy channel and Bob's faulty detectors with perfect ones, use a perfect single-photon source and send no vacuum states.
Under these assumptions, honest Bob will always succeed in measuring the first pulse that Alice sends and disregard the following ones. Hence, Alice's optimal cheating strategy is to create some entangled state, send one qubit to Bob in the first pulse, wait for Bob to reply in step 3 and then perform some measurement in her part of the entangled state in order to decide what to reveal in step 4.  This is no different from the cheating in Berlin et al's protocol, so the optimal cheating probability for Alice is $p_A=\big(3+2\sqrt{a(1-a)}\big)/4$ \cite{berlin}.

{\bf Malicious Bob} -- We consider Bob to be all-powerful, meaning that he controls all aspects of the implementation, except for Alice's photon source. Again, it is in Bob's best interest to replace the lossy and noisy channel with a perfect one, in order to receive each time the correct state and maximize his cheating probability. Moreover, we assume he has perfect detectors and we also give him the ability to know the number of photons in each of the $K$ pulses. Then, Bob's optimal strategy is to receive all $K$ pulses and then perform some operation on the received qubits in order to maximize his information about Alice's bit $c_j$ for some pulse $j$. It is important to note that honest Alice picks a new uniformly random bit $c_j$ for each pulse $j$ and hence Bob cannot combine different pulses in order to increase his information about a bit $c_j$.

To simplify our analysis, we assume that in the case where Bob has received at least two 2-photon pulses or a pulse with 3 or more photons, then he can cheat with probability 1. This probability is in fact very close to 1 and hence our upper bound on Bob's cheating probability is almost tight. We analyze the following events (in each event, the remaining pulses contain zero photons):
\begin{description}
\item[$A_1$]: (all 0-photon pulses) The optimal cheating strategy for Bob is to try to guess Alice's bit, which happens with success probability 1/2.
\item[$A_2$]: (at least one 1-photon pulse) The optimal cheating strategy for Bob is to measure in the computational basis (Helstrom measurement)\cite{helstrom,berlin}. It is proven in \cite{berlin} that this probability is equal to $a$.
\item[$A_3$]: (one 2-photon pulse) It can be proven that for our states the optimal measurement in a 2-photon pulse outputs the correct bit with probability equal to $a$.

\item[$A_4$]: (one 2-photon pulse, at least one 1-photon pulse) Bob will try to benefit from the 2-photon pulse (see discussion below), and if he fails, he will continue like in $A_2$, since the pulses are independent.
\end{description}
Thus, we get an upper bound for the total probability of cheating Bob:
\begin{eqnarray*}
p_B&\leq&\textstyle{\sum_{i=1}^{4}}P(A_i)*P(\text{cheat}\lvert A_i)+(1-\sum_{i=1}^{4}P(A_i) ) \cdot 1
\end{eqnarray*}

Note that an honest Alice prepares the $K$ pulses independently, which means that a measurement on any of them does not affect the rest. Consequently, Bob can measure each pulse independently, without affecting the remaining pulses. Moreover, the probability for each of these events depends on the protocol parameter $K$ and on $\mu$ (which is controlled by Alice).

It remains to bound $P(\text{cheat}\lvert A_4)$, i.e. the case where Bob has received one 2-photon pulse and some single photon pulses. Bob will try to profit from the two identical quantum states in one pulse. On one hand, he can perform the optimal distinguishing measurement on the two photons, which as we said earlier gives a correct answer with probability $a$. On the other hand, he can perform a conclusive measurement on the 2-photon pulse that with some probability will give a correct answer and with some probability will give no answer at all (in which case Bob can use one of the 1-photon pulses). In fact, none of these two strategies is optimal. In general, Bob will perform some measurement that with probability $c$ will provide an answer, which will be correct with probability $\gamma$, and with probability ($1-c$) the measurement will provide no answer, in which case Bob will use a single-photon pulse to guess correctly with probability $a$. Hence,
\begin{center}
$P(\text{cheat}\lvert A_4) = \text{max}_{M} \{c\gamma + (1-c)a\}$
\end{center}
over all possible measurements of Bob.

Let $M$ be the optimal measurement that provides with probability $c$ an answer that is correct with probability $\gamma$ and with probability ($1-c$) provides an answer that is correct with probability $\gamma'$. On one hand, we have that $\gamma'\geq 1/2$ (since Bob can always guess with probability 1/2) and on the other hand this measurement cannot be correct with probability larger than $a$ (since we know that the optimal measurement has probability $a$).
Hence, we have $x \equiv c\gamma+(1-c)1/2 \leq a$ from which we get that $c \geq 2x-1$.
Then, using the above equation we have:
\begin{eqnarray}
P(\text{cheat}\lvert A_4) = c\gamma + (1-c)a \nonumber
                                          &\leq& x+(2-2x)(a-\textstyle{\frac{1}{2}})\nonumber \\
                                          &\leq&-2a^2+4a-1
\end{eqnarray}
Equation (1) provides an analytical upper bound on the cheating probability for event $A_4$ and hence we can now calculate the cheating probability $p_B$.

\COMMENT{{\bf How to choose the measurement basis} --  The most general measurement that Bob can perform on two qubits is in an orthonormal basis of the following form:
\begin{flushleft}
$\lvert\psi_0\rangle=k_{00}\lvert00\rangle+k_{01}\lvert01\rangle+k_{10}\lvert10\rangle+k_{11}\lvert11\rangle\rightarrow 0$\\
$\lvert\psi_1\rangle=l_{00}\lvert00\rangle+l_{01}\lvert01\rangle+l_{10}\lvert10\rangle+l_{11}\lvert11\rangle\rightarrow 1$\\
$\lvert\psi_2\rangle,\lvert\psi_3\rangle \rightarrow X$
\end{flushleft}
The arrows state the bit that he will output in each measurement result, with $X$ being the inconclusive result. Bob is only interested in the basis vectors that provide him with an output bit, i.e. $\lvert\psi_0\rangle$ and $\lvert\psi_1\rangle$. We assume without loss of generality that we have symmetry between the probabilities, meaning that whatever bit Alice chooses and in whatever basis Bob decides to measure it in, the cheating probabilities of all choices should be equal. It can be shown that this leads to $k_{01}=-k_{10}$ and $l_{01}=-l_{10}$. We also have that $c=\big(ak_{00}+(1-a)k_{11}\big)^2 + \big((1-a)k_{00}+ak_{11}\big)^2$ and $\gamma=\frac{1}{c}\big(ak_{00}+(1-a)k_{11}\big)^2$. This means that for each value of $a$ we can find the maximum $p_B$ by maximizing $P(\text{cheat} \lvert A_4)$.}

{\bf Fairness of the protocol} -- In order to have equal cheating probabilities, we adjust $a$ so that $p_A=p_B$.

{\bf Experimental parameters} -- We have introduced a coin flipping protocol that takes into account all experimental parameters. In the following simulations, we use parameter values commonly referenced in the literature \cite{bralut,lut}, which can be implemented using today's technology.

The photon signals that Alice sends arrive with a probability $F$ (transmission efficiency) at Bob's site, and they are detected with a probability $\eta$ (detector quantum efficiency). For an optical channel, $F$ is related to the channel absorption coefficient $\beta$, the channel length $L$ and a distance-independent constant loss $k$, via the equation:
$
F=10^{-(\beta L+k)/10}
$.
The values used in our simulations are shown in the following table.
\begin{center}
\begin{tabular}{l c  c}
\multicolumn{2}{l}{Parameter} & ~~~~~~Value~~~~~~~~ \\
\hline\hline
Receiver constant loss [dB]~~ & ~~~$k$ & 1\\
Absorption coefficient [dB/km]  & ~~~$\beta$  & 0.2 \\
Detection efficiency &~~~$\eta$ & 0.2 \\
Dark counts (per slot) &~~~$d_B$ & $10^{-5}$ \\
Signal error rate &~~~$e$  & 0.01 \\
\end{tabular}
\end{center}

Note that we consider the probability of a signal and a dark count occurring simultaneously negligible.


{\bf Results} -- Our protocol requires a minimum honest abort probability equal to half of the probability of noise in the channel.
We consider an acceptable honest abort probability smaller than 2\%, thus by setting the honest abort probability to fixed values up to 0.02 for different channel lengths, we find the necessary rounds $K$ and the optimal mean photon number $\mu$ that minimize the cheating probability for a fair protocol.

\begin{figure}[!b]
\includegraphics[width = 8.5 cm]{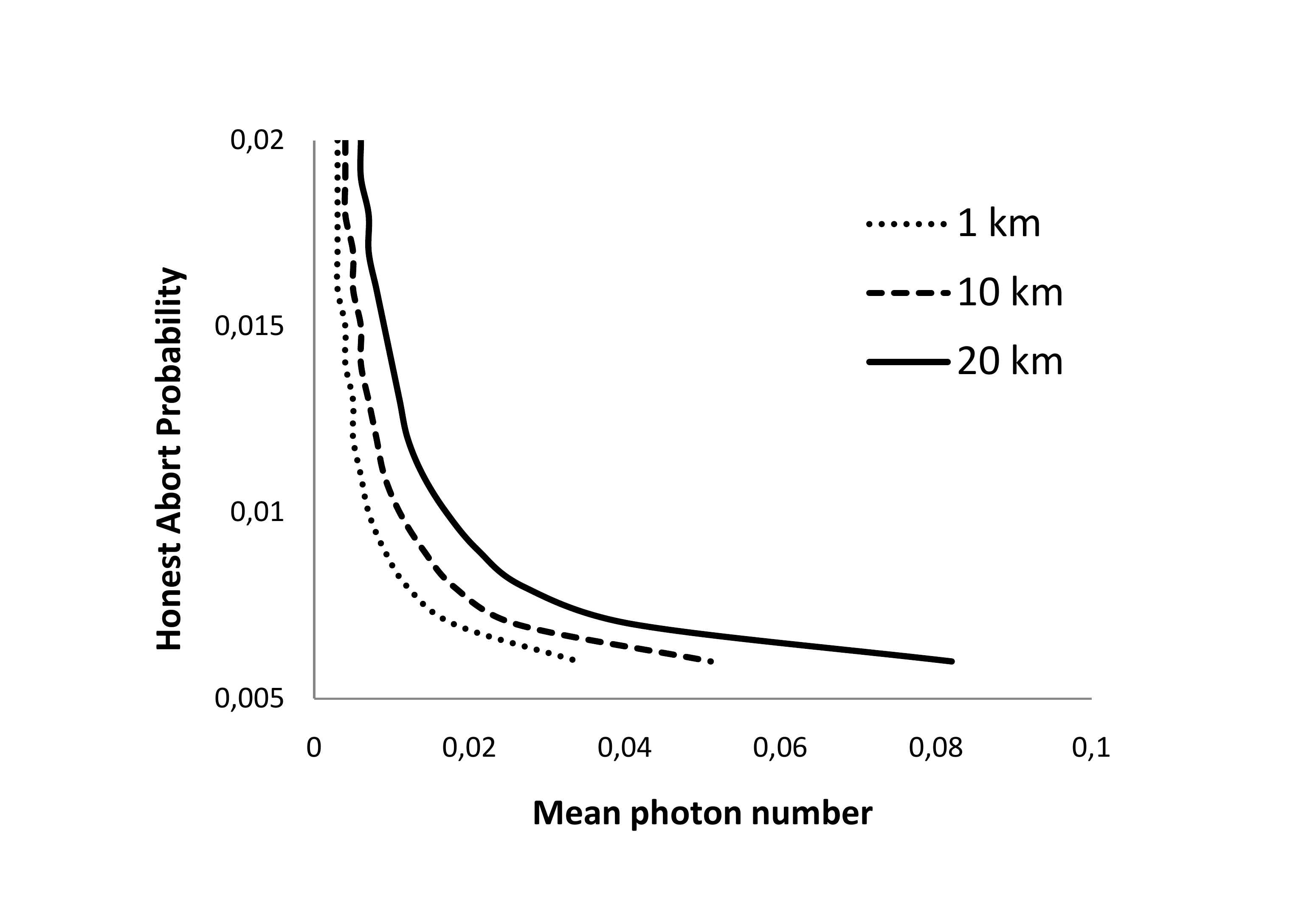}
\caption{Quantum honest abort probability vs mean photon number $\mu$ for different channel lengths.}
\label{fig:abortmean}
\end{figure}

There is an inversely proportional relation between the honest abort probability and the optimal $\mu$ (Figure~\ref{fig:abortmean}).
The same holds for the number of rounds $K$ in relation to the honest abort probability and for the same $\mu$. When $\mu$ is increased in order to achieve the desired honest abort probability, the required number of rounds is reduced. In all our simulations, $K$ did not exceed 15000.

\begin{figure}[!t]
\includegraphics[width = 8.5 cm]{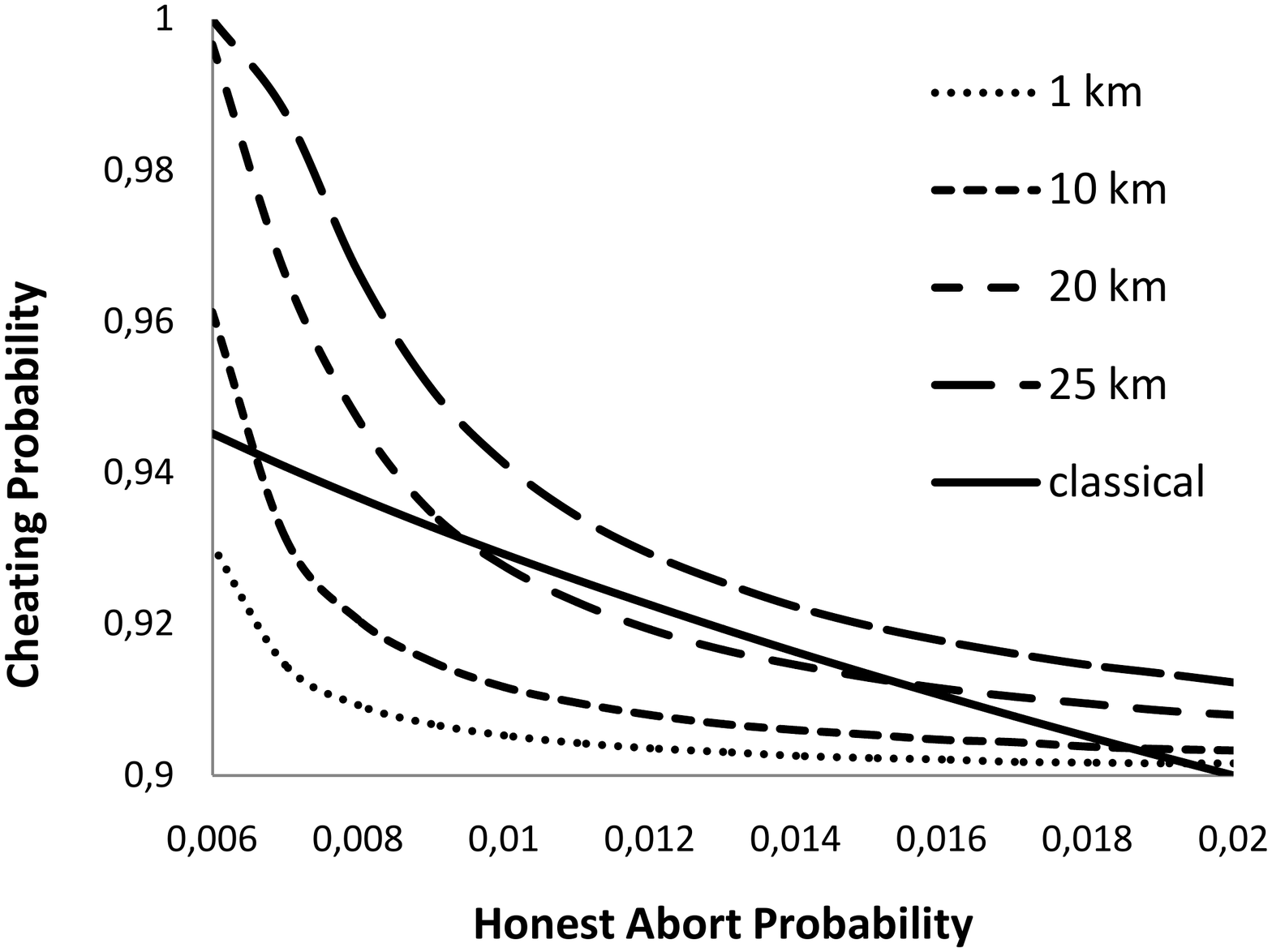}
\caption{Quantum honest abort vs cheating probability for different channel lengths and comparison to the classical case.} \label{fig:abortcheat}
\end{figure}

In Figure \ref{fig:abortcheat} we plot our protocol's cheating probability versus the honest abort probability $H$ for four different channel lengths, and compare this to the optimal classical cheating probability, which is equal to $1-\sqrt{H/2}$ \cite{han}.

\begin{figure}[!t]
\includegraphics[width = 8.5 cm]{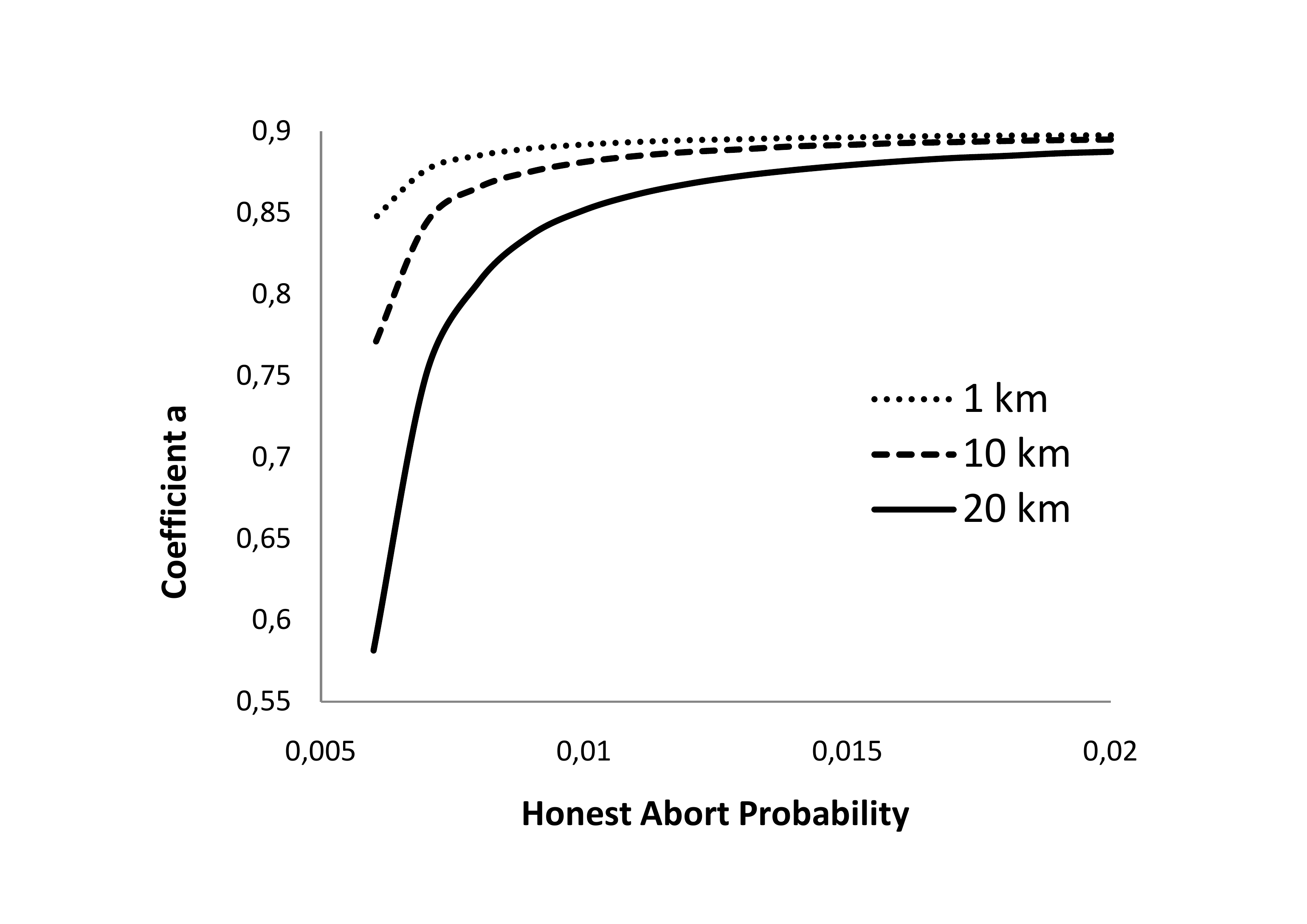}
\caption{Quantum state coefficient vs honest abort probability.}
\label{fig:hon_coef}
\end{figure}

For any length up to 21 km and honest abort probability smaller than 2\%, we can find a $\mu$ such that the maximum cheating probability of our protocol is better than in the classical case. Figure \ref {fig:hon_coef} shows how the coefficient $a$ of the protocol states changes in relation to the honest abort probability, for three different channel lengths.

{\bf Discussion} -- We have shown for the first time that flipping a single coin with security guarantees strictly stronger than in any classical protocol can be achieved with present quantum  technology, and more precisely with a standard attenuated laser source. This implies that quantum information can be used beyond quantum key distribution (QKD), to achieve in practice more difficult cryptographic tasks in a model where the parties do not trust each other. We note that implementations of such tasks will be subject to the same issues related to the existence of side channels as in QKD (eg. \cite{Hack}).

We observe that the maximal communication distance that can be achieved is significantly smaller than in QKD \cite{QKD}. In principle, we cannot expect to have the same results as in QKD, since the setting is much harder. Here, the adversary is the other player, so no cooperation is possible, thus excluding error-correction and privacy amplification. With the parameter values that we used, the limit to the channel length is 21 km. We can increase the channel length by improving the experimental parameters, in particular the signal error rate.

Even though Chailloux \cite{chai10} proposed a protocol with lower cheating probability than the Berlin et al, it does not perform as well in the presence of noise. 

Last, it is interesting to see if there is a way to reduce the effect of noise to the honest abort probability with current technology. We note that this seems hard, since any attempt of Alice to protect the qubits, via a repetition error correcting code for example, will immediately increase the cheating probability of Bob.

{\bf Acknowlegments} -- We acknowledge financial support from the ANR through projects CRYQ (ANR-09-JCJC-0067-01), FREQUENCY (ANR-09-BLAN-0410-01), and QRAC (ANR-08-EMER-012), and from the European Union through project QCS (grant 255961).
\bibliography{CoinFlippingbib}

\end{document}